\documentclass[twocolumn,amsmath,amssymb,aps,prl,superscriptaddress,nofootinbib]{revtex4-2}

\usepackage[english]{babel}
\usepackage[utf8]{inputenc}
\usepackage{xr-hyper}

\usepackage[colorlinks]{hyperref}
\usepackage[T1]{fontenc}
\usepackage[dvips]{graphicx}
\usepackage{amsmath}
\usepackage{amsfonts}
\usepackage{color}
\usepackage{ulem}
\usepackage[caption=false, justification=justified]{subfig}
\usepackage{url}
\usepackage{ulem}
\usepackage[top=2cm, bottom=2cm, left=2cm, right=2cm]{geometry}
\usepackage{setspace}

\usepackage{multirow} 
\usepackage{array}    
\usepackage{booktabs} 
\usepackage[table,xcdraw]{xcolor}  
\usepackage{siunitx}
\usepackage{float}
\usepackage[nolist,nohyperlinks]{acronym}

\makeatletter
\newcommand*{\addFileDependency}[1]{
 \typeout{(#1)}
 \@addtofilelist{#1}
 \IfFileExists{#1}{}{\typeout{No file #1.}}
}
\makeatother

\newcommand{\degree}{$^{\circ}$}

\usepackage{xcolor}

\hypersetup{filecolor=blue}
\hypersetup{citecolor=blue}
\hypersetup{urlcolor=blue} 

\hypersetup{linkcolor=blue}

\begin{document}

\title{Identification of orbital pumping from spin pumping and rectification effects}

\author{Nils Keller} \thanks{These authors contributed equally to this work.}
\affiliation{Institute of Physics, Johannes Gutenberg University Mainz, Staudingerweg 7, 55128 Mainz, Germany}
\affiliation{Department of Applied Physics and Physico-Informatics, Keio University, Yokohama 223-8522, Japan}

\author{Arnab Bose}\thanks{These authors contributed equally to this work.}
\affiliation{Institute of Physics, Johannes Gutenberg University Mainz, Staudingerweg 7, 55128 Mainz, Germany}
\affiliation{Department of Electrical Engineering, Indian Institute of Technology Kanpur, 208016, India}

\author{Nozomi Soya}\thanks{These authors contributed equally to this work.}
\affiliation{Department of Applied Physics and Physico-Informatics, Keio University, Yokohama 223-8522, Japan}

\author{Elias Hauth}
\affiliation{Institute of Physics, Johannes Gutenberg University Mainz, Staudingerweg 7, 55128 Mainz, Germany}
\affiliation{Department of Applied Physics and Physico-Informatics, Keio University, Yokohama 223-8522, Japan}

\author{Fabian Kammerbauer}
\affiliation{Institute of Physics, Johannes Gutenberg University Mainz, Staudingerweg 7, 55128 Mainz, Germany}

\author{Rahul Gupta}
\affiliation{Institute of Physics, Johannes Gutenberg University Mainz, Staudingerweg 7, 55128 Mainz, Germany}

\author{Hiroki Hayashi}
\affiliation{Department of Applied Physics and Physico-Informatics, Keio University, Yokohama 223-8522, Japan}

\author{Hisanobu Kashiki}
\affiliation{Department of Applied Physics and Physico-Informatics, Keio University, Yokohama 223-8522, Japan}

\author{Gerhard Jakob}
\affiliation{Institute of Physics, Johannes Gutenberg University Mainz, Staudingerweg 7, 55128 Mainz, Germany}

\author{Sachin Krishnia}
\email{krishnia@uni-mainz.de}
\affiliation{Institute of Physics, Johannes Gutenberg University Mainz, Staudingerweg 7, 55128 Mainz, Germany}

\author{Kazuya Ando}
\email{ando@appi.keio.ac.jp}
\affiliation{Department of Applied Physics and Physico-Informatics, Keio University, Yokohama 223-8522, Japan}
\affiliation{Keio Institute of Pure and Applied Science (KiPAS), Keio University, Yokohama 223-8522, Japan}
\affiliation{Center for Spintronics Research Network (CSRN), Keio University, Yokohama 223-8522, Japan}

\author{Mathias Kläui}
\email{klaeui@uni-mainz.de}
\affiliation{Institute of Physics, Johannes Gutenberg University Mainz, Staudingerweg 7, 55128 Mainz, Germany}
\affiliation{Graduate School of Excellence Materials Science in Mainz, 55099, Mainz,Germany}
\affiliation{Department of Physics, Center for Quantum Spintronics, Norwegian University of Science and Technology, 7491, Trondheim, Norway}

\date{\today}

\begin{abstract}
The recently predicted mechanism of orbital pumping enables the generation of pure orbital current from a precessing ferromagnet (FM) without the need for electrical current injection. This orbital current can be efficiently injected into an adjacent nonmagnetic material (NM) without being hampered by electrical conductivity mismatch. However, experimentally identifying this novel effect presents significant challenges due to the substantial background contributions from spin pumping and spin rectification effects (SREs). In this work, we disentangle the effects of orbital pumping from spin pumping in bilayer structures composed of Nb/Ni and Nb/$\mathrm{Fe_{60}Co_{20}B_{20}}$ by observing a sign reversal of the measured voltage. This reversal arises from the competing signs of the spin and orbital Hall effects in the Nb. We establish methods to differentiate the pumping signal from SREs by analyzing the distinct angular dependence of the measured voltage and its spatial dependence relative to the radio frequency excitation source.
\end{abstract}

\maketitle

\begin{acronym}[ISHE]

\acro{SHE}{spin Hall effect}
\acro{ISHE}{inverse spin Hall effect}
\acro{STT}{spin transfer torque}
\acro{SOT}{spin-orbit torque}
\acro{SOC}{spin-orbit coupling}
\acro{SHC}{spin Hall conductivity}
\acro{SHA}{spin Hall angle}
\acro{FM}{ferromagnetic}
\acro{NM}{non-magnetic}
\acro{AHE}{anormalous Hall effect}
\acro{SP}{spin pumping}
\acro{FMR}{ferromagnetic resonance}
\acro{OHE}{orbital Hall effect}
\acro{IOHE}{inverse orbital Hall effect}
\acro{OT}{orbital torque}
\acro{OHC}{orbital Hall conductivity}
\acro{MR}{magnetoresistance}
\acro{AMR}{anisotropic magnetoresistance}
\acro{PHE}{planar Hall effect}
\acro{LLG}{Landau-Lifshitz-Gilbert}
\acro{DL}{damping-like}
\acro{FL}{field-like}
\acro{REE}{Rashba-Edelstein effect}
\acro{OREE}{orbital Rashba-Edelstein effect}
\acro{RF}{radio frequency}
\acro{SREs}{spin-rectification effects}
\acro{UV}{ultraviolet}
\acro{OP}{orbital pumping}
\acro{Pt}{platinum}
\acro{Ni}{nickel}
\acro{Nb}{niobium}
\acro{FeCoB}{iron cobald boron}
\acro{Fe}{iron}
\acro{Co}{cobalt}
\acro{Ru}{ruthenium}
\acro{ST-FMR}{spin-torque ferromagnetic resonance}
\acro{DC}{direct current}
\acro{MRAM}{magnetic random-access memory}
\acro{OHA}{orbital Hall angle}
\end{acronym}

\section{Introduction}

Spin and orbital angular momentum are two fundamental properties of electrons, interconnected through \ac{SOC}. In spintronics, the \ac{SOC} is essential in the emergence of various intriguing physical phenomena~\cite{SoumyanarayananNatReview2016} such as stabilization of chiral magentic skyrmions~\cite{Woo2016,Fert2013,Thiaville_2012,Everschor2018} and spin current generation mechanisms~\cite{Manchon2019}. The study of spin currents ($J_\mathrm{S}$), including those generated via the \ac{SHE}~\cite{hirsh1999,sinova2015,miron2011,liu2012Science} and the Rashba-Edelstein effect (REE)~\cite{rashba1983,manchon2015}, has been a central focus in spintronics, especially due to its potential applications in nonvolatile \ac{MRAM}~\cite{BHATTI2017,Dieny2020}.


Recent interest has shifted toward the generation of orbital currents ($J_\mathrm{O}$) due to its potential applications in energy efficient \ac{MRAM} technology ~\cite{gupta2024harnessing, GoEPL}. Theoretical studies suggest that orbital current is a fundamental entity that can for instance result in spin current leveraging the \ac{SOC} of materials ~\cite{GoEPL,GoPRL2018,GoPRB2018}. A major advantage of orbital currents is their potential to be orders of magnitude larger than spin currents across a wide range of materials \cite{Salemi2022}, as they are not inherently limited by the relativistic \ac{SOC}. Consequently, orbital currents could overcome the limitations of spin currents, particularly in terms of scalability and efficiency, making them highly promising for memory and logic applications \cite{GoEPL}.


Thus far, the emerging field of orbitronics has mainly focused on the generation of $J_\mathrm{O}$ through the \ac{OHE} \cite{PRLDing2020,Choi2023,LeeNatCom2021,Kawakami2023,bose2023, ledesma2024non,bose2024fluctuation, hayashi2023observation} and the \ac{OREE} \cite{PRLDing2022,Krishnia2023,Nikolaev2024,ElHamdi2023,Otani2023}. Recently, theorists have predicted the effect of "orbital pumping", where a precessing magnet can emit a significant orbital current without requiring an associated electric current (Figure \ref{devicestructure} (a)) ~\cite{han2023orbital,go2023orbital, ning2024phenomenology}. This effect is analogous to the previously demonstrated spin pumping effect, where a precessing magnet emits pure spin current \cite{brataas2002spin, tserkovnyak2002enhanced, ando2008angular, mosendz2010quantifying, azevedo2011spin, bai2013universal} (Figure \ref{devicestructure} (b)). 

Both spin and orbital pumping provide methods to generate spin and orbital currents without the challenges posed by electrical conductivity mismatch and enable easy and clear detection in comparatively simple samples \cite{ando2011electrically}. As illustrated in Figure \ref{devicestructure} (a,b), the emitted orbital and spin currents are converted into a transverse voltage in the adjacent nonmagnet via reciprocal effects known as the \ac{IOHE}, the inverse orbital Rashba-Edelstein effect \cite{ElHamdi2023, seifert2023time, kumar2023ultrafast, hayashi2024observation} and the \ac{ISHE} and inverse Rashba-Edelstein effect \cite{saitoh2006conversion,Sanchez2013}, respectively. However, in real systems, these processes can occur simultaneously, along with \ac{SREs} arising from the interplay between time-varying magnetoresistance and the applied oscillating field. This overlap makes it highly challenging to isolate the orbital pumping signal from the often dominant background contributions of spin pumping and \ac{SREs} and so methods are needed to identify orbital pumping unambiguously.

In this work, we demonstrate that we can obtain a clear distinction of orbital pumping by carefully selecting materials with opposing signs of the \ac{OHE} and \ac{SHE} and by performing rigorous angular-dependent measurements of the pumped voltage signal. This is achieved using devices specifically designed to produce a more uniform \ac{RF} field while minimizing other parasitic effects as discussed in subsequent sections (Figure \ref{devicestructure} (c)).

\section{Experiment}

\begin{figure*}
    \centering
    \includegraphics[width=\linewidth]{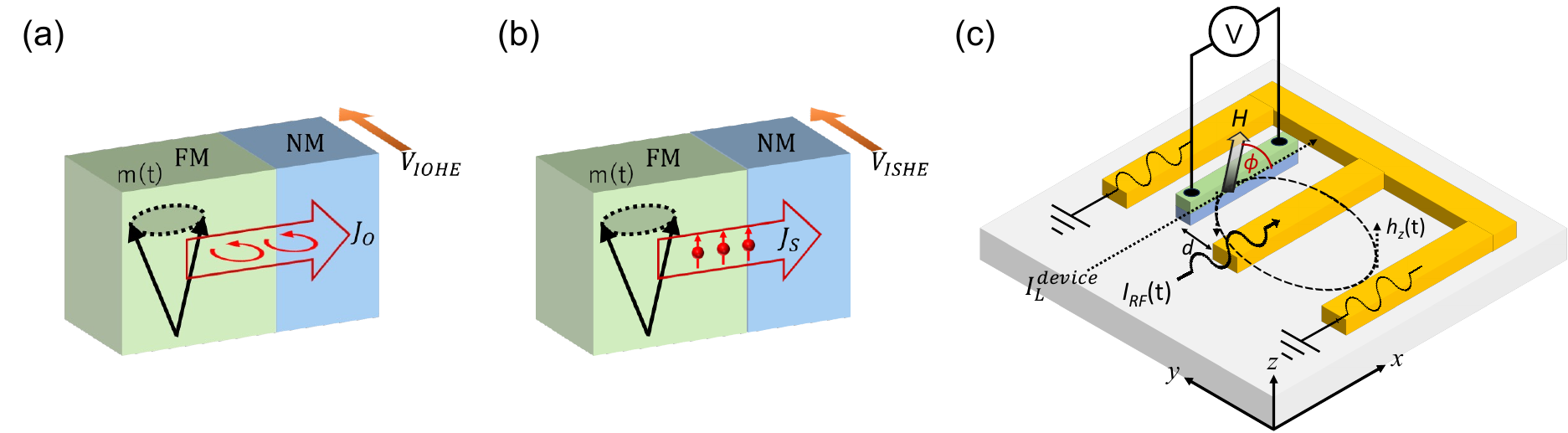}
    \caption{Schematic representation of (a) orbital pumping and (b) spin pumping from a precessing magnet into an adjacent \ac{NM} metal. The orbital and spin currents generated by the precession of the magnet induce a transverse voltage via the  \ac{IOHE} and \ac{ISHE}. (c) Schematic illustration of the device and experimental set up. The \ac{RF} current ($I_\mathrm{{RF}}$) flowing through the waveguide generates an \ac{RF} Oersted field $h_z$ on the \ac{NM}/\ac{FM} device. A fraction of the applied \ac{RF} current is induced in the device ($I_\mathrm{L}^\mathrm{{device}}$) flowing along the longitudinal $x$-direction. Additionally, a static magnetic field $H$ is applied in the plane of the device with an angle $\phi$ with respect to the $x$-axis. The angle convention is chosen such that rotating in the mathematical positive direction increases the value of the angle. The voltage resulting from the orbital and/or spin pumping is measured along the sample wire.}
    \label{devicestructure}
\end{figure*}

We have prepared four primary series of samples using a Singulus Rotaris tool via magnetron sputtering on undoped $\mathrm{Si/SiO_2}$ substrates: (1) Sub/Ta(1)/Nb(4)/Ni(3, 6, 10, 15)/cap, (2) Sub/Ta(1)/Pt(4)/Ni(3, 6, 10, 15)/cap, (3) Sub/Ta(1)/Nb(4)/$\mathrm{Fe_{60}Co_{20}B_{20}(8)}$/cap, and (4) Sub/Ta(1)/Pt(4)/$\mathrm{Fe_{60}Co_{20}B_{20}(8)}$/cap. Here, the numbers in parentheses are the nominal thicknesses in \si{\nano \meter}, and the cap is MgO(1.5)/Ta(1.5) to protect the samples from oxidation. We fabricate the devices shown in Figure \ref{devicestructure} (c) through successive cycles of electron beam lithography, $\mathrm{Ar}^+$ milling, \ac{RF} magnetron sputtering, and lift-off techniques. More information on sample preparation and fabrication is provided in the supplementary information (SI1).

A schematic diagram of the device is shown in Figure \ref{devicestructure} (c). It consists of a coplanar waveguide, with a narrow and long \ac{NM}/\ac{FM} wire placed in the slots of the waveguide with a length of \SI{200}{\micro \meter} and a width of \SI{8}{\micro \meter}. The waveguide and the \ac{NM}/\ac{FM} device, together with the contacts, are electrically isolated by inserting an $\mathrm{SiO_2}$ layer. The central concept is to pass an \ac{RF} current through the waveguide, producing an \ac{RF} magnetic field that drives the magnet into resonance. As a result, the magnet emits both orbital and spin currents into the adjacent NM layer, such as Nb, Pt, or Ru, which are then converted into a \ac{DC} voltage due to the \ac{IOHE} and \ac{ISHE} (Figure \ref{devicestructure} (a,b)).


We chose Nb and Pt as they exhibit opposite signs for the \ac{SHA}, yet have the same sign for the \ac{OHA} \cite{bose2023} from which we can separate the orbital and spin pumping effects via sign reversal. The ferromagnetic materials Ni and FeCoB are selected due to their contrasting abilities to detect orbital currents \cite{LeeNatCom2021, bose2023}. Due to the Onsager reciprocity Ni is therefore expected to generate a pronounced orbital current from the orbital pumping as compared to FeCoB \cite{han2023orbital, go2023orbital}. The Nb- and Pt-based samples are the central focus of this work as they facilitate the unambiguous detection of orbital pumping in our setup via sign reversal in the measured voltage. For comparison, the pumping signal into Ru, another candidate known for exhibiting strong \ac{OHE} and weak \ac{SHE} \cite{bose2023, gupta2024harnessing} is also analyzed in the Sub/Ta(1)/Ru(4)/Ni(10)/cap stack (see supplementary information).


The experiment is performed by sweeping an external magnetic field ($H$) at an angle $\phi$ relative to the $x$-axis, and measuring the voltage in the \ac{NM}/\ac{FM} wire, as shown in Figure \ref{devicestructure} (c). This voltage is fitted using symmetric ($V_\mathrm{S}$) and anti-symmetric ($V_\mathrm{A}$) Lorentzian functions \cite{saitoh2006conversion, mosendz2010quantifying, azevedo2011spin}. In the absence of other \ac{SREs}, $V_\mathrm{S}$ corresponds to the spin/orbital pumping signal, which can be unambiguously identified  via sign reversal in our measurements due to the competing sign of \ac{SHA} and \ac{OHA} in Nb \cite{bose2023}. 


While the spin (and orbital) pumping effects are expected to generate only $V_{\mathrm{S}}$, previous experiments have provided significant evidence for a non-zero value of $V_{\mathrm{A}}$ \cite{saitoh2006conversion, mosendz2010quantifying, azevedo2011spin}. This observation is often attributed to \ac{SREs}, primarily arising from the induced \ac{RF} current within the device (Figure \ref{devicestructure} (c)), which couples with the oscillating magnetoresistance as the \ac{FM} oscillates at ferromagnetic resonance. Consequently, it is critical to determine whether \ac{SREs} also generate $V_{\mathrm{S}}$, as this would complicate the analysis of spin and orbital pumping, potentially leading to incorrect or ambiguous interpretations. One possible origin of $V_{\mathrm{S}}$ is the conventional \ac{ST-FMR} \cite{liu2011spin}, caused by the induced \ac{RF} current flowing through the device. This mechanism, which can produce a significant $V_{\mathrm{S}}$, has largely been overlooked in previous studies \cite{saitoh2006conversion, ando2008angular, mosendz2010quantifying, azevedo2011spin, hayashi2024observation}.

To address this issue, we have fabricated the \ac{NM}/\ac{FM} device within the slot of a waveguide (Figure \ref{devicestructure} (c)). This configuration allows us to distinguish the pumping signal from other \ac{SREs}, which is not feasible when the device is positioned on top of the waveguide, as commonly practiced in earlier studies \cite{harder2016electrical}. In our geometry, the $V_{\mathrm{S}}$ exhibits the following angular dependence as a function of the in-plane magnetic field (\textit{H}) applied at an angle, $\phi$ from the device long axis.
\begin{align}
\label{Symmetricfittingfunction}
    V_\mathrm{S}(\phi) \approx & \, V_\mathrm{S}^{\text{pump}}\sin{\phi} \nonumber + V_{\mathrm{S},\ \text{AMR}}^{\ \text{ST-FMR}}\cos{\phi}\sin{2\phi} \nonumber \\
    & + V_{\mathrm{S},\text{AMR}}^{\text{NL}}\sin{2\phi}
\end{align}

In this work, we primarily focus on the coefficient of $V_\mathrm{S}^{\text{pump}}$ in different samples, which represents the strength of the pumping signal. $V_{\mathrm{S},\ \text{AMR}}^{\ \text{ST-FMR}}$ accounts for contributions from conventional \ac{ST-FMR}, arising from the induced \ac{RF} current and the \ac{AMR} effect as discussed before. $V_{\mathrm{S},\text{AMR}}^{\text{NL}}$ originates from nonlocal \ac{ST-FMR}, where the induced \ac{RF} current couples with out-of-plane magnetization oscillations driven by the Oersted field of the waveguide. Additional minor components of $V_{\mathrm{S}}$, which are negligible in our work, and the origins of $V_{\mathrm{A}}$ are discussed in detail in the supplementary information.





\section{Results and discussion}

\begin{figure*}[htb]
    \centering
    \includegraphics[width=\textwidth]{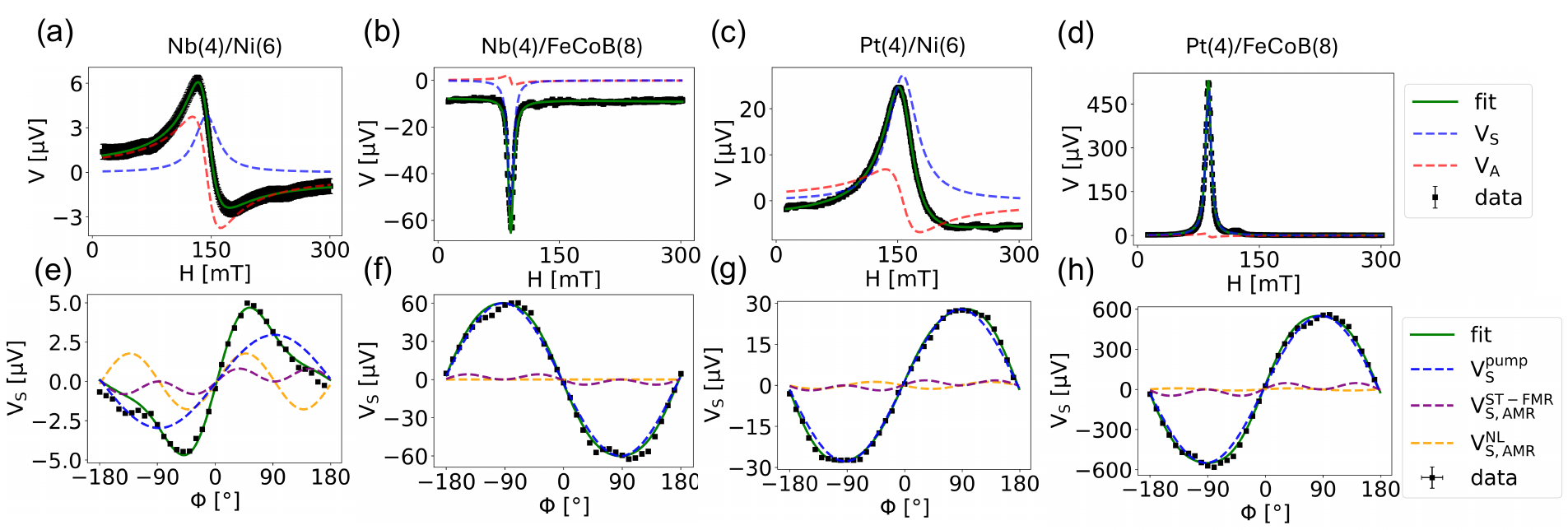}
    \caption{Magnetic field sweep measurements for (a) Nb(4)/Ni(6), (b) Nb(4)/FeCoB(8), (c) Pt(4)/Ni(6) and (d) Pt(4)/FeCoB(8) for a gap width of \SI{6}{\micro \meter} and an angle of \SI{80}{\degree}. The data are fitted with a superposition of symmetric (blue) and antisymmetric (red) Lorentzians. (e-h) The extracted values for \( V_\mathrm{S} \) are plotted as a function of the angle of the swept magnetic field. The data are fitted according to Equation \eqref{Symmetricfittingfunction}.}
    \label{Mainfigure}
\end{figure*}

Figure \ref{Mainfigure} presents the central results of this work. Figures \ref{Mainfigure} (a-d) show the typical voltage spectrum (black squares) in our pumping experiment, fitted with a superposition of symmetric (\( V_\mathrm{S} \)) (blue) and anti-symmetric Lorentzian functions (\( V_\mathrm{A} \)) (red). The fit is displayed in green. The fitting procedure also includes a constant and linear term. The general fitting procedure is discussed in the supplementary information (SI2) in detail. The sign of \( V_\mathrm{S} \) in Pt/Ni (Figure \ref{Mainfigure} (c)) and Pt/FeCoB (Figure \ref{Mainfigure} (d)) reflects the widely studied spin pumping effect in Pt \cite{saitoh2006conversion, ando2008angular, mosendz2010quantifying, azevedo2011spin}. Most interestingly, we observe a sign reversal in \( V_\mathrm{S} \) for Nb/Ni (Figure \ref{Mainfigure} (a)) compared to Nb/FeCoB (Figure \ref{Mainfigure} (b)), which cannot be explained by the conventional spin pumping effect. We find that the sign of \( V_\mathrm{S} \) in Nb/FeCoB (Figure \ref{Mainfigure} (b)) is opposite to that in Pt (Figure \ref{Mainfigure} (c, d)), suggesting that the sign of \ac{ISHE} in Nb is opposite to that in Pt, consistent with theoretical predictions \cite{Salemi2022} and our previous work \cite{bose2023}. In contrast, the observed same sign of \( V_\mathrm{S} \) in Nb/Ni (Figure \ref{Mainfigure} (a)) and Pt/(Ni or FeCoB) (Figures \ref{Mainfigure} (c,d)) strongly suggests that the orbital pumping effect dominates in the Nb/Ni samples where the injected orbital current is converted into an electrical voltage via the \ac{IOHE}.



Next, we analyze the influence of other \ac{SREs} on our measurements by studying the angular dependence of $V_\mathrm{S}$, as shown in Figures \ref{Mainfigure} (e-h). We observe that \( V_\mathrm{S} \) can be reasonably well fitted with Equation \eqref{Symmetricfittingfunction}, indicating a prominent contribution from spin and orbital pumping (\( \sin{\phi} \) term) along with non-zero values of other \ac{SREs} across all samples. These findings confirm that the sign of the \ac{SHA} is negative for Nb and it is positive for Pt (compare Figures \ref{Mainfigure} (f, g, h)). Further, the angular dependence reveals that the pumping signal (\( \sin{\phi} \) term in $V_\mathrm{S}$) in Nb/Ni is positive (\ref{Mainfigure} (e)), indeed predominantly driven by orbital pumping prevailing over spin pumping. 

Our work demonstrates that \ac{SREs} are stronger in Ni-based samples than in FeCoB films, as Ni exhibits the highest \ac{AMR} among other transition metal magnets (Co, Fe). This enhanced \ac{AMR} leads to more significant \ac{SREs} from the undesired spin and orbital currents, and various current-induced magnetic fields (more details are provided in the supplementary information SI3). The sign reversal of the $\sin{\phi}$ term, clearly, allows us to detect orbital pumping via the \ac{IOHE}. We also conducted an analogous experiment in Ru/Ni (see supplementary information SI4) that shows large signals from orbital pumping due to the vanishingly small \ac{SHA} and the predicted large \ac{OHA} \cite{bose2023}.

\begin{figure}
    \centering
    \includegraphics[width=\linewidth]{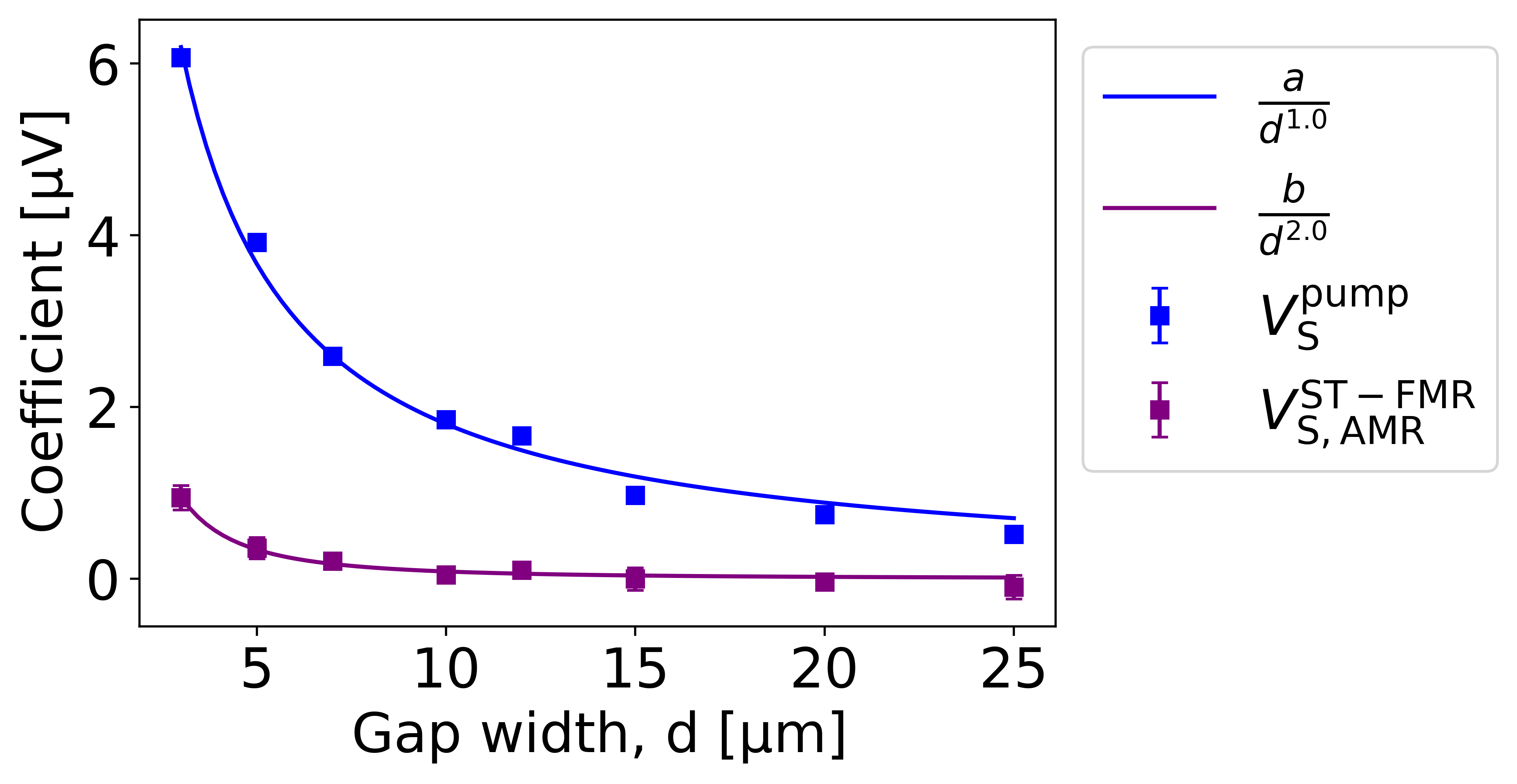}
    \caption{The strengths of the different effects in Nb(4)/Ni(10) bilayers are characterized by the coefficients from Equation \eqref{Symmetricfittingfunction}. The coefficients are plotted as a function of the gap width, $d$. The $V_\mathrm{S}^\mathrm{pump}$ data is fitted with $\frac{a}{d^m}$ obtaining a value of $m \approx 1.0$ for the best fit.}
    \label{Figure3_Gapdependence}
\end{figure}


Building upon these observations, we investigate the dependence of the pumping effect and \ac{SREs} on the gap width ($d$) between the device and the waveguide in Nb(4)/Ni(10) samples (Figure \ref{Figure3_Gapdependence}). The individual data points are obtained using Equation \eqref{Symmetricfittingfunction}. We find that the pumping signal ($V_\mathrm{S}^\mathrm{{pump}}$, blue curve in Figure \ref{Figure3_Gapdependence}) goes down as $d$ increases and is proportional to $\frac{1}{d^{m}}$ with $m \approx 1.0$ in our work. Since the spin-pumping voltage is proportional to the power of the microwave magnetic field, it should scale with $\frac{1}{d^2}$. However, due to the finite size of the waveguide and the device a behaviour of $m < 2$ is expected. The voltage component arising from $V_\mathrm{{S,AMR}}^\mathrm{{ST-FMR}}$ also follows a similar trend (purple squares in Figure \ref{Figure3_Gapdependence}) but exhibits a steeper decrease with $m \approx 2.0$. Such a steep decrease in $V_\mathrm{{S,AMR}}^\mathrm{{ST-FMR}}$ is expected as it results from the interplay between the induced \ac{RF} current and its Oersted field within the device, both of which diminish as the spacing increases. We find a similar behaviour in the bilayer system Ru(4)/Ni(10) (see supplementary information SI5). Thus, our findings demonstrate that the spacing dependence is an efficient way to separate the pumping signal from other \ac{SREs}.


\section{Summary and Conclusion}

In this work, we systematically investigate spin and orbital pumping effects in Nb/Ni and Nb/FeCoB bilayers, comparing our findings with those from Pt/Ni and Pt/FeCoB systems. We examine the DC voltage spectrum generated by orbital and spin pumping, along with other rectification effects, under uniform microwave excitation, ensured by positioning the lithographically fabricated bilayer device within the waveguide slot. Our methodology includes analyzing the angular dependence of the measured voltage signal as a function of the in-plane magnetic field (Figure \ref{Mainfigure}), as well as its variation with the separation between the waveguide and the device (Figure \ref{Figure3_Gapdependence}). This approach enables us to uniquely distinguish the pumping signal from undesired additional rectification effect signal contributions. We confirm the orbital pumping effect by observing the sign reversal of the $\sin \phi$ component of the symmetric Lorentzian voltage signal in Nb/Ni compared to Nb/FeCoB and its spacing dependence, which provides a distinct signature from spin rectification effects. Thus, we not only demonstrate an efficient method to generate orbital current via the "orbital pumping" mechanism from a precessing magnet but also establish a robust approach to disentangle it from spin pumping and other rectification effects.


\section{Acknowledgement}
The authors thank the DFG (Spin+X (A01, A11, B02) TRR 173-268565370 and Project No. 358671374), the Horizon 2020 Framework Programme of the European Commission under FETOpen Grant Agreement No. 863155 (s-Nebula); the European Research Council Grant Agreement No. 856538 (3D MAGiC); and the Research Council of Norway through its Centers of Excellence funding scheme, Project No. 262633 ”QuSpin”. The study also
has been supported by the European Horizon Europe Framework Programme under an EC Grant Agreement N°101129641 ”OBELIX”. AB acknowledges the support from the Alexander von Humboldt Foundation for his postdoctoral fellowship. KA acknowledges the support from JSPS KAKENHI (Grant No. 22H04964), Spintronics Research Network of Japan (Spin-RNJ), and MEXT Initiative to Establish Next-generation Novel Integrated Circuits Centers (X-NICS) (Grant No. JPJ011438).

\subsection{Author contributions}
The samples were grown by FK, RG and GJ and the devices were fabricated by NK and NS with inputs of AB. NK and EH carried out the experiments and analyzed the data with inputs from AB who conceived the idea. SK and GJ assisted in the experiments and also in data analysis. HK and HH assisted in theory. The manuscript was written by AB, NK, and SK. The whole project was supervised by KA and MK. All authors commented on the manuscript.

\subsection{Data Availability}
Data is made available from the corresponding author upon reasonable request.

\bibliographystyle{ieeetr}
\bibliography{references}

\end{document}